\documentclass[%
 aip,
 amsmath,amssymb,
 reprint,%
]{revtex4-1}

\usepackage{graphicx}% Include figure files
\usepackage{dcolumn}% Align table columns on decimal point
\usepackage{bm}% bold math
\usepackage[utf8]{inputenc}
\usepackage[T1]{fontenc}
\usepackage{mathptmx}
\usepackage{etoolbox}
\usepackage{xcolor}
\usepackage{tabularx}
\usepackage{booktabs}

\makeatletter
\def\@email#1#2{%
 \endgroup
 \patchcmd{\titleblock@produce}
  {\frontmatter@RRAPformat}
  {\frontmatter@RRAPformat{\produce@RRAP{*#1\href{mailto:#2}{#2}}}\frontmatter@RRAPformat}
  {}{}
}%
\makeatother
\begin{document}

\preprint{AIP/123-QED}

\title{Direct data-driven forecast of local turbulent heat flux in Rayleigh-B\'{e}nard convection}
% Force line breaks with \\
\author{Sandeep Pandey}
\affiliation{Institute of Thermodynamics and Fluid Mechanics, Technische Universit{\"a}t Ilmenau, D-98684 Ilmenau, Germany.}

\author{Philipp Teutsch}
\affiliation{Institute for Practical Computer Science and Media Informatics, Technische Universit{\"a}t Ilmenau, D-98684 Ilmenau, Germany.}

\author{Patrick M\"ader}
\affiliation{Institute for Practical Computer Science and Media Informatics, Technische Universit{\"a}t Ilmenau, D-98684 Ilmenau, Germany.}
\affiliation{Faculty of Biological Sciences, Friedrich-Schiller-Universit\"at Jena, D-07745 Jena, Germany.}

\author{J\"org Schumacher}
\affiliation{Institute of Thermodynamics and Fluid Mechanics, Technische Universit{\"a}t Ilmenau, D-98684 Ilmenau, Germany.}
\affiliation{Tandon School of Engineering, New York University, New York, NY 11201, USA.}

\date{\today}

\begin{abstract}
A combined convolutional autoencoder--recurrent neural network machine learning model is presented to analyse and forecast the dynamics and low-order statistics of the local convective heat flux field in a two-dimensional turbulent Rayleigh-B\'{e}nard convection flow at Prandtl number ${\rm Pr}=7$ and Rayleigh number ${\rm Ra}=10^7$. Two recurrent neural networks are applied for the temporal advancement of flow data in the reduced latent data space, a reservoir computing model in the form of an echo state network and a recurrent gated unit. Thereby, the present work exploits the modular combination of three different machine learning algorithms to build a fully data-driven and reduced model for the dynamics of the turbulent heat transfer in a complex thermally driven flow. The convolutional autoencoder with 12 hidden layers is able to reduce the dimensionality of the turbulence data to about 0.2 \% of their original size. Our results indicate a fairly good accuracy in the first- and second-order statistics of the convective heat flux. The algorithm is also able to reproduce the intermittent plume-mixing dynamics at the upper edges of the thermal boundary layers with some deviations. The same holds for the probability density function of the local convective heat flux with differences in the far tails. Furthermore, we demonstrate the noise resilience of the framework which suggests the present model might be applicable as a reduced dynamical model that delivers transport fluxes and their variations to the coarse grid cells of larger-scale computational models, such as global circulation models for the atmosphere and ocean.  
\end{abstract}

\maketitle

\section{\label{sec:intro} Introduction}

Turbulent thermal convection processes form one fundamental class of flows that are found in numerous natural and technological applications ranging from astrophysical scales in stellar interiors to sub-meter lengths in heat exchangers. \cite{Kadanoff2001,Ahlers2009,chilla2012new,SchumacherRMP2020} The fundamental physical question in these flows is the one on the local and global mechanisms of turbulent heat transfer which is typically significantly enhanced by the turbulent fluid motion in comparison to a purely diffusive transport in a quiescent medium. In its simplest configuration, a thermal convection flow consists of a fluid layer which is enclosed by two impermeable parallel plates at distance $H$, known as the Rayleigh-B\'{e}nard convection case. The bottom plate is uniformly heated at a constant temperature $T=T_0+\Delta T$ and the top plate is cooled at $T=T_0$.\cite{Verma2018} For temperature differences $\Delta T>0$ being large enough, the buoyancy-triggered fluid motion is turbulent. Convective turbulence is sustained by characteristic coherent structures which are denoted as {\em thermal plumes}. These unstable fragments of the thermal boundary layer permanently rise from the bottom or fall from the top into the bulk region of the convection layer and thus inject kinetic energy into the flow. Thermal plumes are also the local building blocks of the global heat transfer; their morphology has been studied in several experimental \cite{Zhou2007,Moller2021} and numerical studies.\cite{Shishkina2008,Emran2012} They are connected with the local convective heat flux which is given by 
\begin{equation}
	j_{\rm conv}({\bm x},t)=u_z({\bm x},t) \theta({\bm x},t)\,,
	\label{conv1}
\end{equation} 
with 
\begin{equation}
	\theta({\bm x},t)=T({\bm x},t)-\langle T(z)\rangle_{A,t}\,.
	\label{conv2}
\end{equation} 
Here, $u_z$ is the vertical velocity component and $\theta$ the deviation of the total temperature field $T$ from the mean profile $\langle T(z)\rangle_{A,t}$. The analysis of this flux requires the joint solution of the coupled Boussinesq equations for the velocity and temperature fields. Here, we want to model this central transport quantity and its statistical properties directly without solving the underlying nonlinear equations by means of recurrent neural networks. This results in a significant simplification and data reduction and sets the major motivation for the present work. 

Machine learning (ML) methods have caused a change of paradigms to analyse, model and control turbulent flows. \cite{Jordan2015,LeCun2015,Kutz2017,Duraisamy2019,BrennerPRF2019,Brunton2020,Pandey2020} This evolution is driven by the growing technological capabilities of numerical and laboratory experiments to generate high-dimensional, highly resolved data records at increasing Reynolds or Rayleigh number that can reproduce many aspects of fully developed turbulent flows in great detail. Flow features, such as the thermal plumes in the present case, have then to be classified, dynamically modeled, or connected to statistical moments for parametrizations and other reduced descriptions. To illustrate the typically resulting demands of a data analysis better, let us consider a concrete example of a three-dimensional direct numerical simulation (DNS) of a turbulent thermal convection flow. A layer of height $H$ with an aspect ratio of $60H:60H:H$ at a Rayleigh number $Ra\sim 10^8$ is resolved with about 6 billion spectral collocation points on an unstructured spectral element mesh for each of the involved fields, such as the three velocity components, temperature, and pressure.\cite{Vieweg2021} It amounts to 181 GByte of raw data per snapshot. The estimate does not incorporate the temporal dynamics that is typically stored as a sequence of such highly resolved snapshots. This underlines clearly the necessity to process and reduce data in completely new ways to uncover the main physical processes, such as the characteristic structures that form the backbone of the turbulent heat transfer.     

Reduced-order models (ROM) are derived to approximate the dynamics of the most energetic degrees of freedom or the large-scale flow and predict low-order turbulence statistics such as mean or fluctuation profiles. Most of these models are data-driven and can be generated in several ways, e.g., by Proper Orthogonal Decomposition (POD) \cite{lumley1967structure,berkooz1993proper}, Dynamic Mode Decomposition \cite{schmid2010dynamic}, nonlinear Laplacian spectral analysis \cite{GiannakisMajda2012} or expansions in modes and eigenfunctions of the Koopman operator \cite{Rowley2009,Giannakis2018} to mention only a few. Particularly, the POD is still a popular workhorse for projection-based reduced models \cite{Moehlis_et_al_2002, noack_papas_monkewitz_2005, bailon2011, soucasse_podvin_riviere_soufiani_2020} and has been combined more recently also with ML algorithms.\cite{pawar2019deep, renganathan2020machine, deng2019time, rahman2019nonintrusive} With the increase of the vigor of turbulence (which is in line with an increase of Reynolds or Rayleigh number), the number of necessary POD modes in a ROM grows quickly. As a consequence, limitations of these models are reached quickly even with efficient algorithms such as the snapshot method.\cite{Sirovich1987} This circumstance calls for alternative ways to reduce simulation snapshots which is a further motivation for our present work.

In the present work, we combine a convolutional autoencoder (CAE) with recurrent neural networks (RNNs) to obtain a ML-based equation-free dynamical model for the convective heat flux field $j_{\rm conv}({\bm x},t)$. The convolutional encoder reduces the high-dimensional simulation snapshots of the convective heat flux to a low-dimensional feature space. In this latent space, the RNNs are trained and then run autonomously to advance the dynamics of $j_{\rm conv}({\bm x},t)$ with respect to time. A subsequent convolutional decoder transforms the resulting latent space data back into high-dimensional data snapshots of the flux. The choice of the hyperparameters of the RNNs is discussed in detail. We discuss two CAE-RNN architectures which will predict the low-order statistics, such as the mean and fluctuation profiles of the convective heat flux, very well and are even able to reproduce the probability density function (PDF) of $j_{\rm conv}({\bm x},t)$. The two chosen RNN architectures are as follows:  
\begin{itemize}
\item {\em Echo state networks} (ESNs) belong to the class of reservoir computing models \cite{jaeger2004ESN} and have been used to model the nonlinear dynamics of the Lorenz 63 or Lorenz 96 models, chaotic acoustic models, and two-dimensional convection without and with phase changes.\cite{lu2017reservoir,vlachas2020backpropagation,Huhn2022,Pandey2020a,Heyder2021} In the latter cases, the ESN was combined  with a data reduction by POD \cite{Pandey2020a,Heyder2021}, such that the temporal dynamics of the POD expansion coefficients was modeled. 
\item {\em Gated Recurrent Units} (GRUs) with a specific sequence to sequence (seq2seq) architecture form the second RNN architecture that is studied in the present work. More precisely, we apply an encoder-decoder GRU \cite{cho2014learning} that was originally designed for natural language processing, but is now also applied to predict continuous time series data.\cite{du2018time,sangiorgio2020robustness} 
\end{itemize}
Here, we thus substitute the data reduction and expansion that was formerly done by means of a snapshot POD \cite{Pandey2020a,Heyder2021} by a convolutional encoder/decoder network, which will be better suited for higher-dimensional simulation data. Differently to previous studies, we feed the derived field $j_{\rm conv}({\bm x},t)$ directly into the ML algorithm.  Throughout this work, we will remain in the two-dimensional Rayleigh-B\'{e}nard convection setup to demonstrate our concepts. For example in ref. \cite{Pandey2020a}, we were able to compress our data by 92\% while loosing 17\% of the turbulent variance. This is improved here even further to a compression by 99.7\% while loosing only about 5\% of the variance.
%------------------------------------------------------
\begin{figure*}[htb]
	\includegraphics[width=1.0\linewidth]{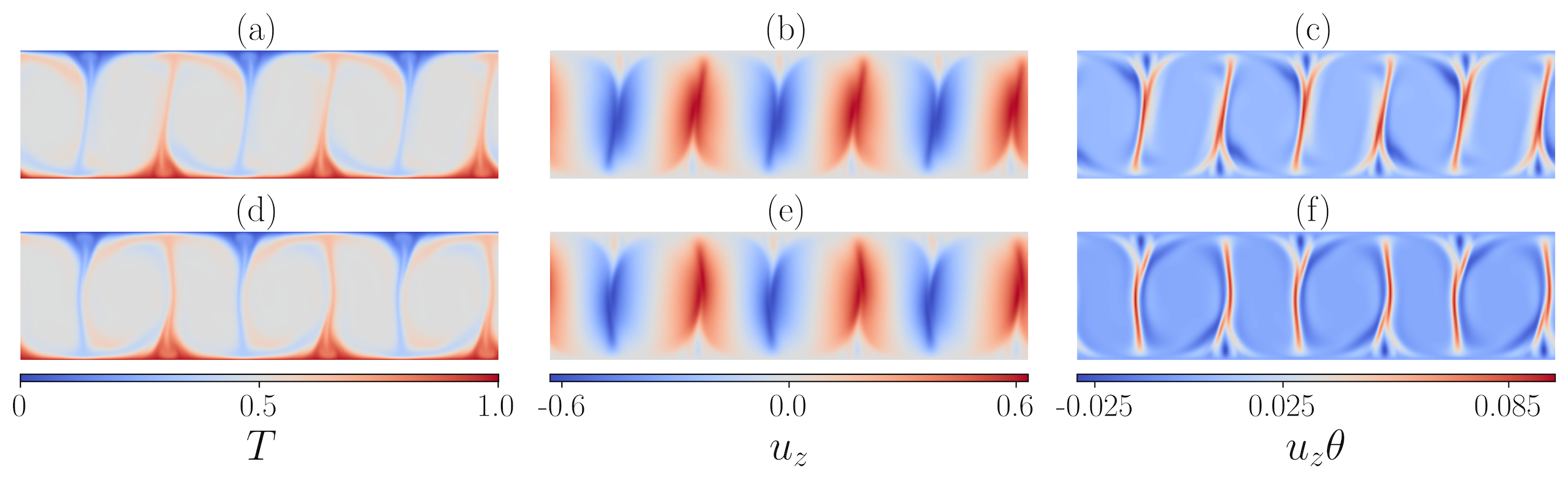}
	\caption{Contours of the two-dimensional convection fields for two time instances. (a,d) Total temperature $T$, see also eq. \eqref{conv2}. (b,d) Vertical velocity component $u_z$. (c,f) Resulting turbulent convective heat flux $j_{\rm conv}$ as given by eq. \eqref{conv1}.}
	\label{Figure00}
\end{figure*} 
%------------------------------------------------------

Finally, we mention that RNN-based encoder-decoder architectures are often used as comparison model for ESN models in similar applications. \cite{al2017review,han2019review,bianchi2020reservoir,qian2021attention} Encoders and decoders have been successfully employed for de-noising data \cite{lu2013speech, gondara2016medical}, anomaly detection \cite{zhou2017anomaly} and to dimensionality reduction \cite{hinton2006reducing, wang2016auto} in many other fields. Encoder and decoder networks can incorporate a variety of nonlinear activation functions, thus taking advantage of higher-order representations in connection with a deep network architecture. Particularly for multi-dimensional datasets, such architectures can reduce the training effort due to its parameter sharing and sparse connectivity. \cite{Goodfellow2016} For example, the spatio-temporal dynamics of fluid flows past cylinders and airfoils \cite{xu2020multi,omata2019novel,murata2020nonlinear} or for turbulent channel flows \cite{fukami2020convolutional} were successfully predicted and analyzed with CAEs. It should be mentioned that Koopman methods have been recently combined with RNNs.\cite{Eivazi2021} Neural network algorithms have been also used to fit the global laws of turbulent heat and momentum transfer better.\cite{Bhattacharya2022}

The outline of the manuscript is as follows. Section II presents the Boussinesq equations of turbulent convection and the DNS in brief that generate the data base. Section III discusses in detail the building blocks of our CAE-ESN and CAE-GRU networks including the hyperparameter tuning. Section IV discusses the training procedures and the results of both model runs with test data. We summarize the work and give a brief outlook at the end in section V. Further specific details on the architecture of the CAE and the training are summarized in appendices A and B, respectively.

\section{Simulation Data of Two-Dimensional Turbulent Convection}
\label{DNS}
The turbulent convection data are generated by a DNS using nek5000 spectral element solver \cite{FISCHER199784} in the two-dimensional case. The Boussinesq equations \eqref{eq:NS1}--\eqref{eq:NS3}, which couple the velocity components $(u_x, u_z)$ and temperature $T$, are solved in a closed rectangular cell of an aspect ratio $L/H=6$. They are given in dimensionless form by
%----------------------------------------------------------
\begin{align}
	\frac{\partial u_i}{\partial x_i} &= 0 \,,
	\label{eq:NS1}\\
	\frac{\partial u_i}{\partial t} + {u_j} \frac{\partial u_i}{\partial x_j} &= - \frac{\partial p}{\partial x_i} + \sqrt{\frac{\text{Pr}}{\text{Ra}}} \,\frac{\partial^2 u_i}{\partial x_j^2} + T \delta_{i,3} \,,
	\label{eq:NS2}\\
	\frac{\partial T}{\partial t} + {u_j} \frac{\partial T}{\partial x_j} &= \frac{1}{\sqrt{\text{RaPr}}} \,\frac{\partial^2 T}{\partial x_j^2}\,.
	\label{eq:NS3}
\end{align}
%----------------------------------------------------------
The pressure field is denoted by $p$ and $i,j \in \{x,z\}$. The horizontal coordinate is given by $x$, the vertical one by $z$. The dimensionless Rayleigh number Ra is a measure of the vigor of convective turbulence,  set to ${\rm Ra}=10^7$ here. The dimensionless Prandtl number Pr which is the ratio of momentum to thermal diffusion was fixed to ${\rm Pr}=7$, as for thermal convection in water. All equations are made dimensionless by the cell height $H$, free-fall velocity $U_f = \sqrt{g\alpha \Delta TH}$ ($g$ is the acceleration due to the gravity and $\alpha$ the thermal expansion coefficient) and the temperature difference between the bottom and top plates $\Delta T >0$. The simulation apply $48\times 16$ spectral elements. On each element, the 4 fields are expanded in polynomials of order 11 in each space dimension. For the machine learning analysis, these fields are interpolated spectrally on a uniform grid consisting of $N_x\times N_z=320\times 60$ points. As a result, we have 2400 snapshots and thus a total of $2400\times 320\times 60$ data points. Snapshots were sampled at every $0.125$ free-fall time units, $H/U_f$. This data generation process using the DNS took approximately 123 CPU-hours including the initialization with random perturbations. More details on the simulation and the boundary conditions can be found in ref. \cite{Pandey2020a}. 

Figure~\ref{Figure00} shows snapshots of the two-dimensional convection fields. We display the temperature (left column) and vertical velocity component (middle column) that are combined to the derived convective heat flux field $j_{\rm conv}$ (right column). The latter field is characterized by sharp ridges which suggest a strongly localized convective heat flux. Exactly these ridges will be extracted as the dominant features by the CAE. We note that the Nusselt number, a global dimensionless measure of the turbulent heat transfer, follows ${\rm Nu}=1+\sqrt{\rm Ra Pr}\langle j_{\rm conv}\rangle_{A,t}\ge 1$. The symbol $\langle \cdot\rangle_{A,t}$ stands for a combined average with respect to time $t$ and area  $A=L\times H$.

\section{Building Blocks of the End-to-End Pipeline}
\begin{figure*}[htb]
	\includegraphics[width=0.7\textwidth]{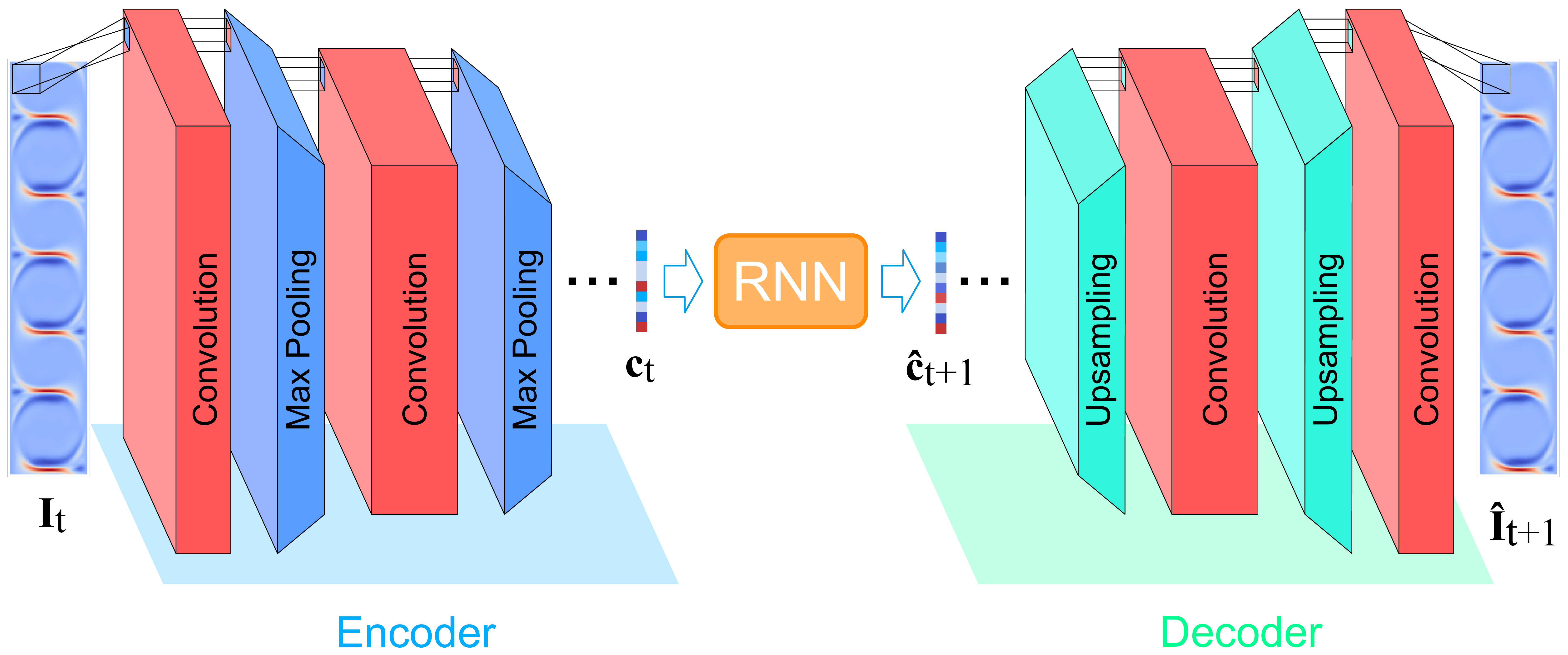}
	\caption{Illustration of proposed end-to-end pipeline for the forecasting of convective heat flux dynamics. The convolutional autoencoder receives the high-dimensional field data of the direct numerical simulations at time $t$ and has been trained for a task-specific order reduction. A trained RNN consumes the reduced order dynamics compressed by the encoder and forecasts future dynamics in the reduced order space at time $t+1$. The compressed dynamics is input for the decoder that accompanies the CAE to predict a fully resolved flow field at time $t+1$. Note that the RNN can also be run in latent space for several time steps $p>1$ from time $t$ to $t+p$.}
	\label{Figure01}
\end{figure*}

Figure~\ref{Figure01} illustrates the building blocks and workflow in our end-to-end pipeline. It consists of a convolutional encoder for compressing the two-dimensional spatio-temporal data, a recurrent neural network (RNN) to forecast dynamics in the reduced order space (also known as latent space), and a accompanying convolutional decoder decompressing the two-dimensional convective heat flux field from the forecasted reduced order data. We study two types of RNNs for forecasting, an ESN and a GRU-based network. Once trained, the presented purely data-driven approach can process simulation snapshots on the fly. In contrast, the previously applied POD snapshot analysis \cite{Pandey2020a,Heyder2021} requires knowledge of the entire training dataset ab initio to extract the POD modes from the collection of simulation snapshots and thus to obtain subsequently the time series of the POD expansion coefficients as the RNN input. The following subsections discuss the individual building blocks of our CAE--RNN approach in detail.     
%------------------------------------------------------
%------------------------------------------------------

\subsection{Generation of compressed snapshot representation}\label{sec:3}
An autoencoder is a machine learning model consisting of an encoder and a decoder part, see Fig.~\ref{Figure01}. The purpose of the encoder is to compress its input into a trained latent space. The accompanying decoder takes data in this latent representation and  reconstructs its representation in the original input domain. From a learning perspective, autoencoders are self-supervised, i.e., the network's input is also used as expected output and no additional labelling effort for a training dataset is required. That is, in theory an autoencoder encodes input data ${\bm I}$ into ${\bm c} = {\rm encode}({\bm I})$ where ${\bm c} \in \mathbb{R}^{l}$. It decodes ${\bm c}$ back into $\hat{{\bm I}} = {\rm decode}({\bm c})$ subsequently. In practice, however, there will be deviations, such that $\hat{{\bm I}}\approx {\bm I}$ (cf. Fig.~\ref{Figure01}). An $L_2$ norm is used as an objective function and the training uses a gradient descent method, e.g., with adaptive momentum (Adam).\cite{kingma2014adam}

A CAE utilizing convolutional layers rather than fully connected layers is specifically suitable for coping with the complexity of high-dimensional input data. A typical convolutional layer combines a convolution operation on the input data with trainable kernels and an activation function to induce a nonlinearity. Consider, three-dimensional input data ${\bm I} \in \mathbb{R}^{C_{\rm in}\times N_x\times N_z}$ with $C_{\rm in}$ the number of input channels. Here, $C_{\rm in}=1$ and thus ${\bm I} \in \mathbb{R}^{N_x\times N_z}$. This input is convoluted with a kernel $k_m$ as shown in Eq.~\eqref{eq:cnn}, where $m\in [1, C_{\rm out}]$ with the number of output channels $C_{\rm out}$. Furthermore, $b_m$ is the bias term of kernel $k_m$, and $\psi$ a nonlinear activation function. Zero-padding is typically applied to a layer's input to prevent information loss at the edges of the input data,
%------------------------------------------------------
\begin{equation} 
    {\rm Conv}(m, {\bm I}) = \psi \left(\sum_{i=1}^{C_{\rm in}} k_m * {\bm I}_i \right) +b_m\, \text{ for } m \in [1, C_{\rm out}] \,.
    \label{eq:cnn}
\end{equation}
%------------------------------------------------------
In the encoder, one or multiple convolutional layers are typically followed by a pooling layer using a window of configurable size and sliding with a configurable step across the spatial dimensions of the layer's input. Thereby, the input per window is aggregated into a single output element with a selectable aggregation function. This aggregation function of the common max-pooling layer retains only the maximum element per window. The pooling layer's step size is typically chosen so that it reduces the dimensionality of input ${\bm I}$. The eventual output of the encoder is the input data represented in latent space $\mathbb{R}^{l}$ where $l\ll N_x \times N_z$. The CAE's decoder follows a mostly analogous design. The main difference is the use of upsampling layers rather then max-pooling layers to decode data in latent space back to the input domain shape. The effective processing of complex turbulence data requires an architecture of multiple convolutional and accompanying upsampling layers in order to gain a compressed representation without substantial loss of information (cf. Table~\ref{tab:cae} in appendix A). In the end-to-end scenario and after training all networks, we utilize the CAE's encoder to derive a compressed form ${\bm c}_t$ from a given snapshot ${\bm I}_t$. The compressed snapshot becomes input to a subsequent RNN that forecasts the next reduced representation $\hat{\bm c}_{t+1}$, which is then decoded back into the input domain by the CAE's decoder.

\subsection{Forecast of compressed snapshots}
We study two types of RNNs for forecasting compressed snapshots in the latent domain, echo state networks (ESN) and gated recurrent units (GRU).

\subsubsection{Echo state network}\label{sec:ESN}
\begin{figure}[htb]
	\includegraphics[width=0.9\linewidth]{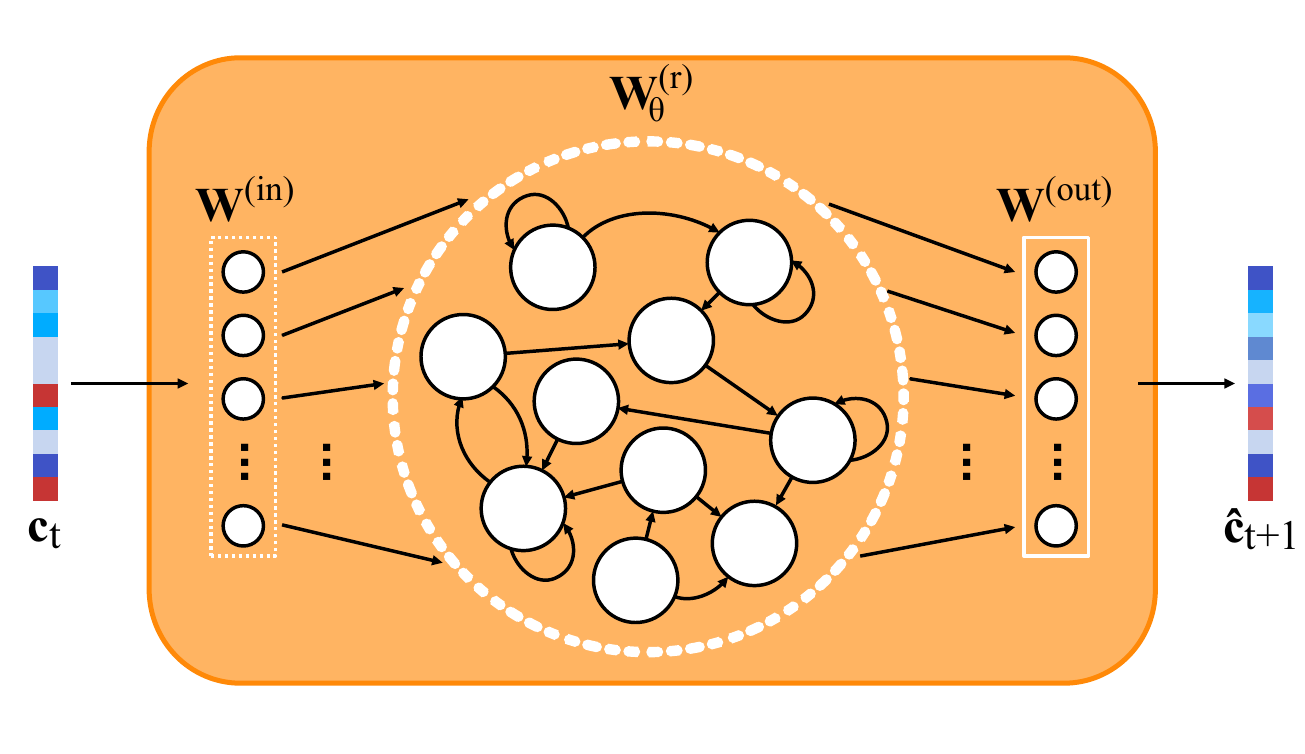}
	\includegraphics[width=0.9\linewidth]{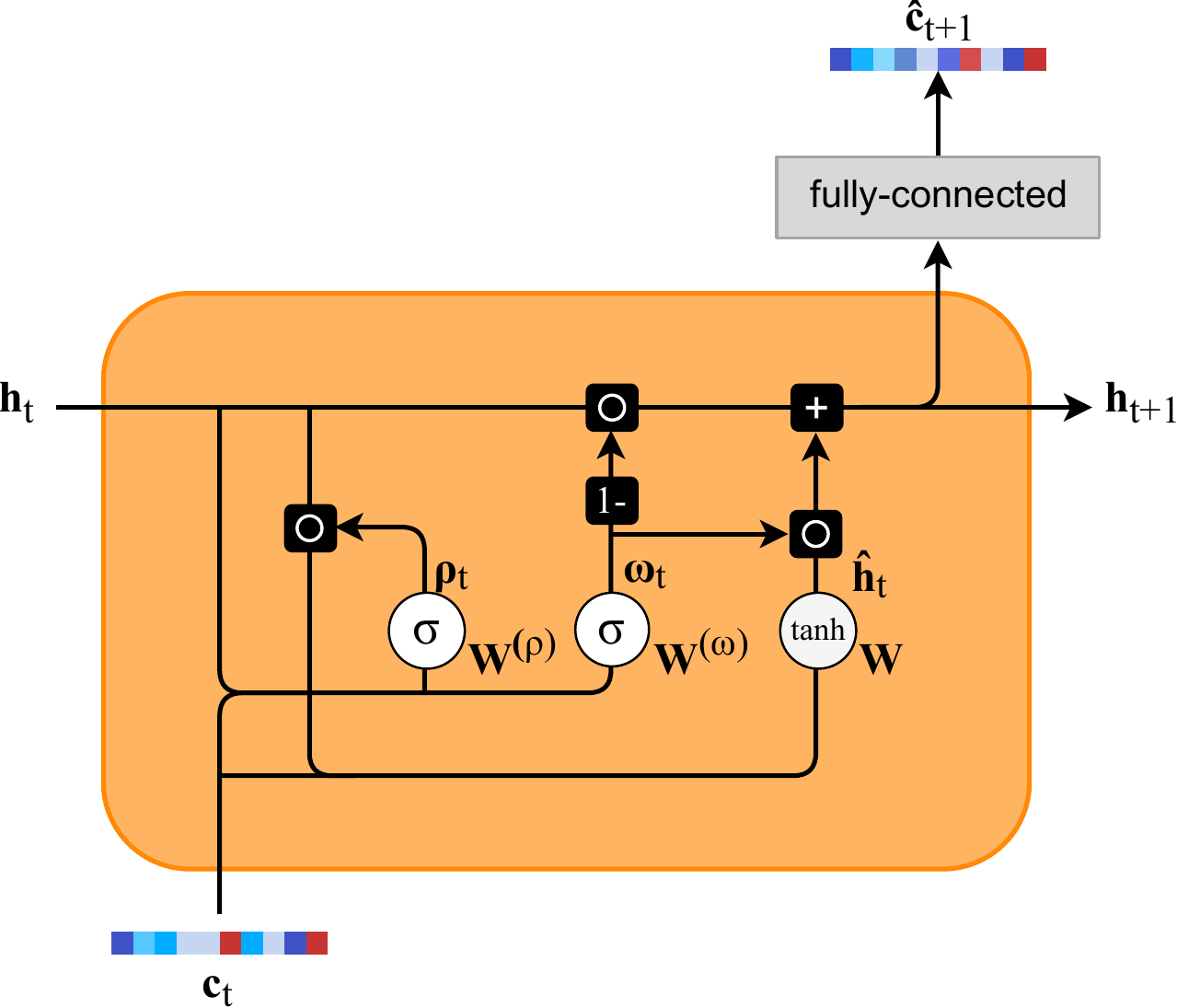}
	\caption{Top: Echo state network (ESN) architecture consisting of an input layer, the reservoir, and an output layer. Bottom: Gated recurrent unit (GRU) cell consisting of a reset gate and an update gate.}
	\label{fig:esnCell}
\end{figure}

The ESN has previously shown great potential in modeling sequential turbulent flow data.\cite{vlachas2020backpropagation,chattopadhyay2020data} ESNs consist of an input layer; a sparsely occupied and randomly parametrized network of recurrent and direct connections, the reservoir; and an output layer, see Fig.~\ref{fig:esnCell} (top). 
%------------------------------------------------------
%------------------------------------------------------
\begin{figure*}[htb]
	\includegraphics[width=1.0\linewidth]{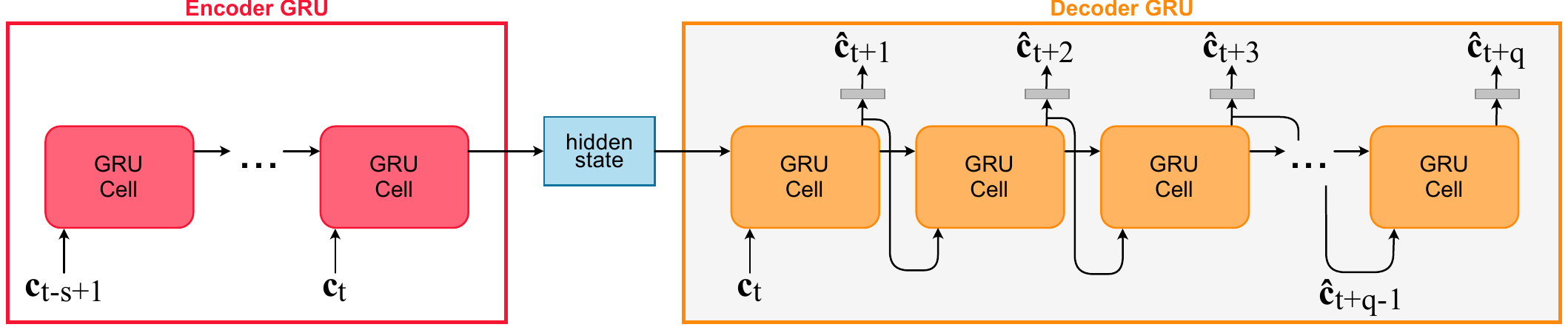}
	\caption{Forward data flow of an encoder-decoder network based on a gated recurrent unit.}
	\label{fig:encDec}
\end{figure*} 
%------------------------------------------------------

The reservoir is represented as an adjacency matrix $\mathbf{W}_\eta^{(r)}$ whose initialization depends on a vector of hyperparameters $\eta$ (cf. Section~4) and is used to encode the ESN's input into a hidden representation, the reservoir, that accumulates information of previous inputs. This reservoir is furthermore updated per time step with new input. More specifically, at each time step $t$ the ESN's input ${\bm c}_t$ influences the computation of an updated reservoir state ${\bm r}_{t}\in \mathbb{R}^N$ which is computed by 
%------------------------------------------------------
\begin{equation}
{\bm r}_{t} = (1-\alpha) {\bm r}_{t-1} + \alpha \cdot tanh\left(\mathbf{W}^{({\rm in})} {\bm c}_{t} + \mathbf{W}_\eta^{(r)}{\bm r}_{t-1}\right),
\label{RC1}
\end{equation}
%------------------------------------------------------
where $N$ denotes the size of the reservoir with $N\gg M_x\times M_z$ by ${\bm r}_t=\mathbf{W}^{({\rm in})} {\bm c_{t}}$ and $\alpha$ denotes a leakage rate determining the blending of previous state and current input. The random matrices $\mathbf{W}^{({\rm in})}$ and $\mathbf{W}_\eta^{(r)}$ are initialized at the beginning of the training and remain unchanged thereafter. The ESN's output at $t$ is obtained by
\begin{equation}
	\hat{\bm c}_{t+1}=\mathbf{W}^{({\rm out})} {\bm r}_{t}\,, 
\end{equation}
where the components of $\mathbf{W}^{({\rm out})}$ are the only trained parameters, optimized by a mean squared error cost function with a ridge regression regularization in a non-iterative manner. The hyperparameters of the ESN are the reservoir size $N$, the node density $D$ which is the precentage of active nodes, the spectral radius of the reservoir $\rho(\mathbf{W}_\theta^{(r)})$, the leakage rate $\alpha$, and the Tikhonov regularization parameter $\beta$ in the cost function. Thus $\eta=(N,D,\rho,\alpha,\beta)$. This training is inherently fast compared to other types of RNN. \cite{vlachas2020backpropagation,jaeger2004ESN} After successful training and hyper-parameter optimization, the ESN runs as an autonomous dynamical system, i.e., a forecasted compressed snapshot ${\bm c}_{t+1}$ can be used as the next ESN input to forecast ${\bm c}_{t+2}$ and so on. 

\subsubsection{Gated recurrent unit}
As already mentioned in the introduction, we also study a gated recurrent unit (GRU) in an encoder-decoder architecture as an RNN, used for complex sequence analysis and forecasting problems. GRU is an advanced RNN cell that uses gates to control which information becomes part of the maintained cell state and which previously acquired information can be forgotten. Due to this mechanism, the GRU effectively mitigates vanishing and exploding gradient problems typically faced when training  RNNs with long training sequences. Figure~\ref{fig:esnCell} (bottom) shows the interplay of the components of the GRU.

The operation of the reset gate and the update gate is denoted as 
%------------------------------------------------------
\begin{equation}
    {\bm \rho}_t = \sigma({\bm W}^{(\rho)} \cdot [{\bm h}_{t},{\bm c}_t])\,,
\end{equation}
%------------------------------------------------------
and
%------------------------------------------------------
\begin{equation}
    {\bm \omega}_t = \sigma({\bm W}^{(\omega)} \cdot [{\bm h}_{t},{\bm c}_t])\,,
\end{equation}
%------------------------------------------------------
where ${\bm \rho}_t$ denotes the reset gate vector and ${\bm \omega}_t$ the update gate vector at time step $t$; ${\bm W}^{(\rho)}$ and ${\bm W}^{(\omega)}$ are the corresponding weight matrices applied to a vector formed by concatenating the input vector ${\bm c}_t$ at time step $t$ with the hidden state vector ${\bm h}_{t}$. The sigmoid activation function $\sigma$ ensures an output range between $0$ and $1$. The output is used in an element-wise vector multiplication to determine how much of other vector element's value to preserve. More specifically, ${\bm \rho}_t$ is element-wise multiplied with ${\bm h}_{t}$ to ``reset'' individual values of the old state when computing the updated intermediate state $\hat{\bm h}_t$ as 
\begin{equation}
    \hat{{\bm h}}_t = \tanh({\bm W} \cdot [{\bm \rho}_t \circ {\bm h}_{t},{\bm c}_t])\,,
\end{equation}
where ${\bm W}$ is an additional weight matrix and $\tanh$ is used as activation function. Analogously, ${\bm \omega}_t$ is element-wise multiplied with ${\bm h}_{t}$ and $(1-{\bm \omega}_t)$ with $\hat{\bm h}_t$ to add current input information to the updated cell state ${\bm h}_{t+1}$, which is also the cell's output, given by 
\begin{equation}
    {\bm h}_{t+1} = (1-{\bm \omega}_t) \circ {\bm h}_{t} + {\bm \omega}_t \circ \hat{{\bm h}}_t\,.
\end{equation}
The output is then fed forward through a final fully-connected layer with a linear activation to forecast the next compressed snapshot $\hat{\bm c}_{t+1}$.

We organize two GRU cells in an encoder-decoder architecture  \cite{bahdanau2014neural,chung2014empirical} as detailed in Fig.~\ref{fig:encDec}. The encoder processes a sequence of compressed input snapshots while building the hidden state that represents a latent representation thereof. This latent representation is then passed to the decoder, which uses the encoded accumulated information up to the previous time step to forecast the next compressed snapshot of the sequence, $\hat{c}_{t+1}$. Following an initial externally triggered step, the decoder progresses auto-regressively; it consumes its output of the previous iteration $\hat{c}_{t+1}$ as an input for the computation of an updated latent representation which is used to forecast the next output $\hat{c}_{t+2}$.

\section{Results and discussion}
Before we turn to the training, validation, and test phases, we list a few more details on the DNS data record. In total, we used 2,400 snapshots divided into three subsets. The first subset consisted of 1,000 snapshots exclusively used for training, thus, called training set. The second subset, called validation set, consisted of 500 snapshots used to cross-validate the training and to warrant the robustness and generalization ability of a given model in the runtime. The third subset contained the remaining 900 snapshots used as an independent, unseen, and non-trained test data set. We trained CAE and GRU on NVIDIA GeForce GTX 1060 and RTX 2080TI GPUs, respectively, and the ESN on a CPU with 16 GBytes of memory.

\subsection{Training of the convolutional autoencoder}
\begin{figure}[htb]
\centering
\includegraphics[width=\linewidth]{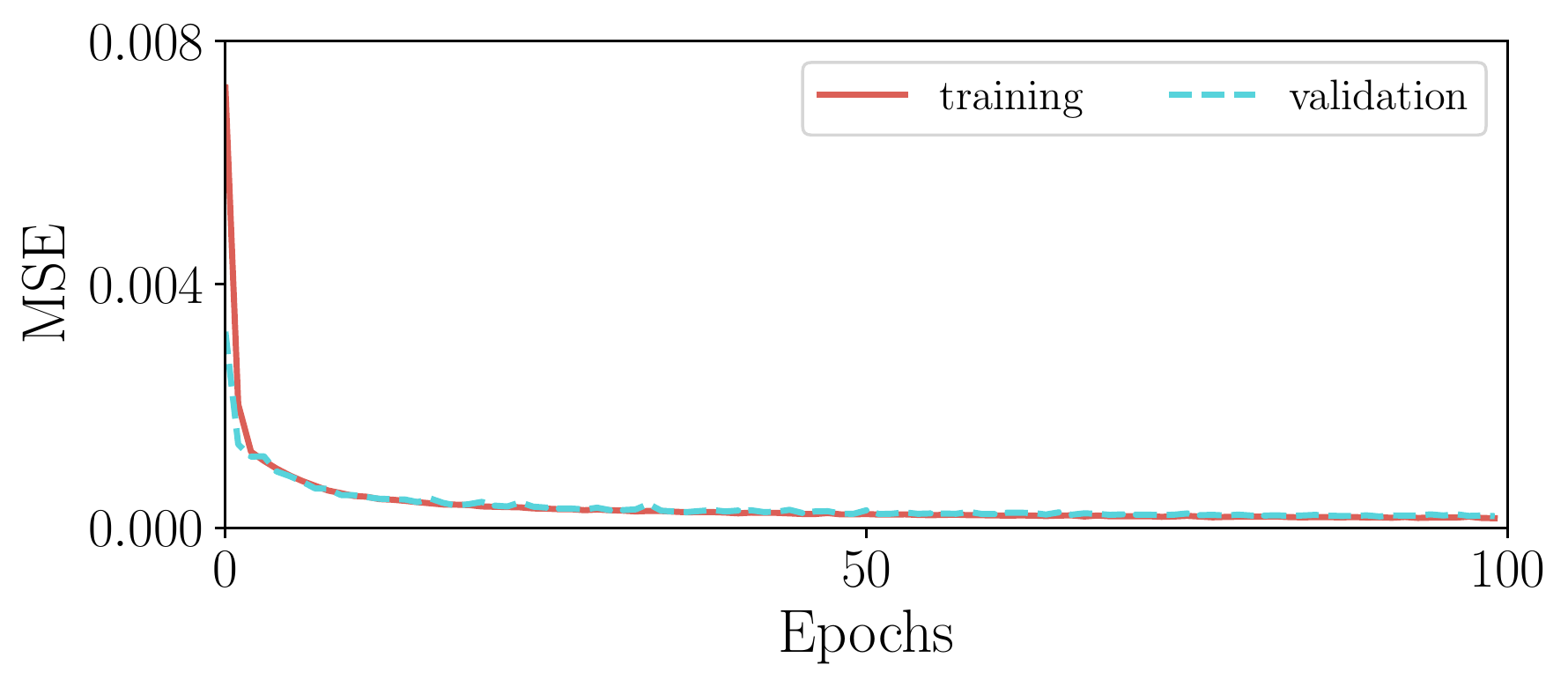}
\caption{The loss function in the form of a mean squared error (MSE) versus the number of training epochs.}
\label{Figure02}
\end{figure}

We first developed a multi-layer CAE consisting of 12 layers for each, encoder and decoder (cf. Tab.~\ref{tab:cae} in appendix~A). The decoder consists of an additional cropping layer to gain the original data snapshot dimension of $320\times 60$. This final layer uses a sigmoid activation to ensure a normalized output. We optimized the network hyperparameters using a Bayesian optimization (BO). Table~\ref{tab2} shows the optimized parameters, the value search range per parameter, and the eventually discovered optimum within this range. Eventually, we used convolutions with a kernel size of $5 \times 5$ giving them a larger receptive field. For the latent representation between encoder and decoder, we found a size of 40 elements to best suit our application. The latent space will thus be 40-dimensional. All network weights were initialized following a Glorot uniform distribution.\cite{glorot2010understanding} More details on the BO procedure are detailed in appendix B. After obtaining the optimized parameters, we trained the CAE and simultaneously cross-validated the model against the validation dataset at the end of every epoch (cf. Fig.~\ref{Figure02}). The training has converged after about 30 epochs; it is however continued until it satisfies the early stopping criterion. In addition, we observe that both training and validation error decrease coherently which indicates the robustness of the model on unseen validation data. 

%------------------------------------------------------
%----------------------------------------------------
\begin{table}[htb]
\begin{tabular}{lcc} 
\hline\hline                            
Parameter & Search range & Optimized value \cr
\hline
Kernel size & ($1\times 1$)--($5\times 5$)  & $5\times 5$  \cr
Latent vector size & $\big\{20k\mid k\in\{1,2,\ldots,3\}\big\}$  & 40 \cr
Learning rate & 0.0001--0.001  & 0.00058  \cr
Batch size & $\big\{8l\mid l\in\{1,2,3\}\big\}$  & 16  \cr
\hline\hline
\end{tabular}
\caption{Optimized parameters obtained from the Bayesian optimization for the convolutional autoencoder. The optimization process was started with 5 random initial points and thereafter 25 iterations were used with a factor $\kappa=1$, see eq. \eqref{ucb} in appendix B.}
\label{tab2}
\end{table}
%------------------------------------------------------

\subsection{Training of the echo state network}
After obtaining a trained CAE delivering compressed representations of input data, we proceeded with the training of the ESN for the prediction of the temporal evolution of the convection flow in the latent space. We employed the same dataset and the same splitting as discussed above. We used the mean squared error (MSE) between the predicted and ground truth modes as an objective function for ESN training. We also monitored the MSE between the predicted and the original turbulent convective heat flux as an additional metric for training success. This was realized by continuously feeding the predicted test modes to the decoder and gather them in ensembles. Again, we optimized the network's hyperparameter using a BO. 

\begin{figure}[htb]
\includegraphics[width=\linewidth]{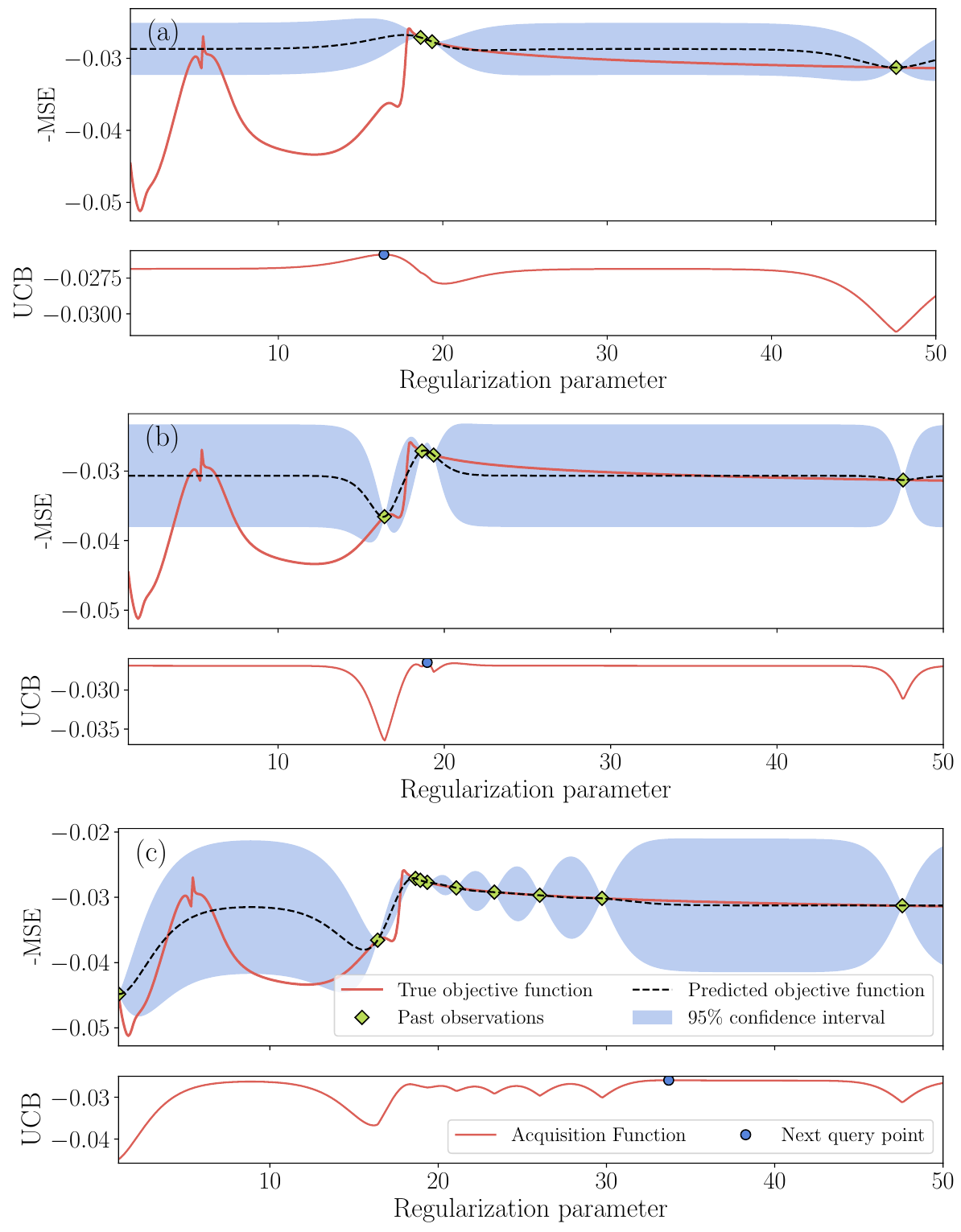}
\caption{Exemplary progress of the Bayesian optimization of the ESN. In panels (a) to (c), we plot the negative mean squared error (MSE) in the top together with the upper confidence bound in panel (c) for the search of the optimal Tikhonov regularization parameter $\beta$. (a) Iteration No.~1 after initialization with 2 random points is shown. (b) Iteration No.~2 is shown. (c) Iteration No.~8 is shown. For this demonstrative case, we took $\kappa=1$ (see appendix B), and the following ESN hyperparameters: $\alpha = 0.96$, $\rho= 0.92$, $D= 0.2$, and $N = 100$ from the given ranges.}
\label{Figure03}
\end{figure}

Figure~\ref{Figure03} illustrates an example of the BO progress for three different iterations. In the example, we solely optimize the regularization parameter $\beta$ while keeping the other hyperparameters constant. In these figures, the actual (unknown, black-box) objective function is shown as the red curve. One can observe the complex nature along with an incapability of a grid-search if the grid is too coarse to capture the optima. Here, we used the MSE as a cost function and maximized the negative of MSE. 

We started the BO at 2 random points for $\beta$ which enable the calculation of the posterior distribution as shown in Fig. \ref{Figure03}(a). The third observation is at $\beta=18.6$ and the acquisition function in the form of an upper confidence bound (UCB) predicts the next query point at $\beta=16.4$ because it becomes the area for exploitation as $\kappa=1$. In Figure~\ref{Figure03}(b), it can be seen that the uncertainty becomes zero for $\beta=16.4$ (assuming no noise) and that the acquisition function suggests a next point, $\beta=18.9$. This iteration proceeds until we reach an optimum or a predefined number of iterations. Figure \ref{Figure03}(c) illustrates the status after 8 iterations, showing that the model is still not converged, but continues to explore a region with higher uncertainty and eventually yielding an optimum. Here, the factor $\kappa=1$ in the UCB is taken which forces the algorithm to exploit regions with higher mean, see again appendix B. In our BO for the ESN, we optimized the 5 hyperparameters in vector $\eta$ plus the scaling. They are known to have a significant effect on the performance of the network. Table~\ref{tab:BO_ESN} summarizes the optimized hyperparameters obtained after 50 BO iterations. Due to the high-dimensionality of our input data, the ESN is prone to overfitting which is visible in the cross-validation phase. To mitigate this problem, we added a negative penalty term to the cost function.

%------------------------------------------------------
%------------------------------------------------------
\begin{table}[htb]
\begin{tabular}{lcc} 
\hline\hline                              
Parameter & $\quad$Search range$\quad$ & Optimized value \cr 
\hline
Reservoir size & 100--5,000  & 2,992  \cr
Spectral radius & 0.90--0.99  & 0.97  \cr
Reservoir density & 0.05--0.20  & 0.09  \cr
Scaling & true, false & false  \cr
Leakage rate & 0.5--0.9   & 0.50  \cr
Regularization parameter& 0--600  & 4.89  \cr
\hline\hline
\end{tabular}
\caption{Optimized parameters obtained from BO for the ESN. Here, 5 points were randomly chosen to initialize the prior, 50 iterations were used for the BO and $\kappa=1$.} 
\label{tab:BO_ESN}
\end{table}

\subsection{Training of the gated recurrent unit}
We further went on to train the encoder-decoder GRU as an alternative to the ESN. Both will predict the temporal evolution of the convection flow in the latent space. Again, we searched for the optimal hyperparameters of the network in our given application scenario. More specifically, we optimized here the learning rate, the batch size, and the hidden state size via a grid-search (cf. Tab.~\ref{tab:gru_hyperparameters}). Differently to the ESN, the GRU is trained by means of a stochastic gradient descent (SGD) method, equally to the CAE. Therefore, we use snapshots from the training, validation and test set respectively to generate samples. Each sample consists of $50$ input snapshots and $100$ target snapshots. Within the training process we dynamically adapt the learning rate once the training loss did not improve for a given number of epochs (patience), i.e., we used a reduce learning rate on plateau scheduler. The reduction is performed by multiplying the current learning rate with a factor $\gamma$. We set the number of patience epochs to $20$ and $\gamma = 0.6$. Finally we applied an early stopping to determine the number of training epochs. Here we choose a patience of $100$ epochs to ensure the convergence of the model.

\begin{table}[htb]
    \centering
    \begin{tabular}{lcc}
    	\hline\hline
         Parameter & Search range & Optimized value  \\
         \hline
         Initial learning rate & 0.006, 0.003, 0.001, & 0.001\\
                               & 0.0006, 0.0003, 0.0001, &\\
                               & 0.00006, 0.00003, 0.00001 &\\
         Batch size & 32, 64, 128, 256, 512 & 128 \\
         Hidden state size & 128, 256, 384, 512, 1024 & 512 \\
         \hline\hline
    \end{tabular}
    \caption{Hyperparameters used for the encoder-decoder GRU training}
    \label{tab:gru_hyperparameters}
\end{table}

%------------------------------------------------------
\begin{figure*}[htb]
\centering
\includegraphics[width=1.0\linewidth]{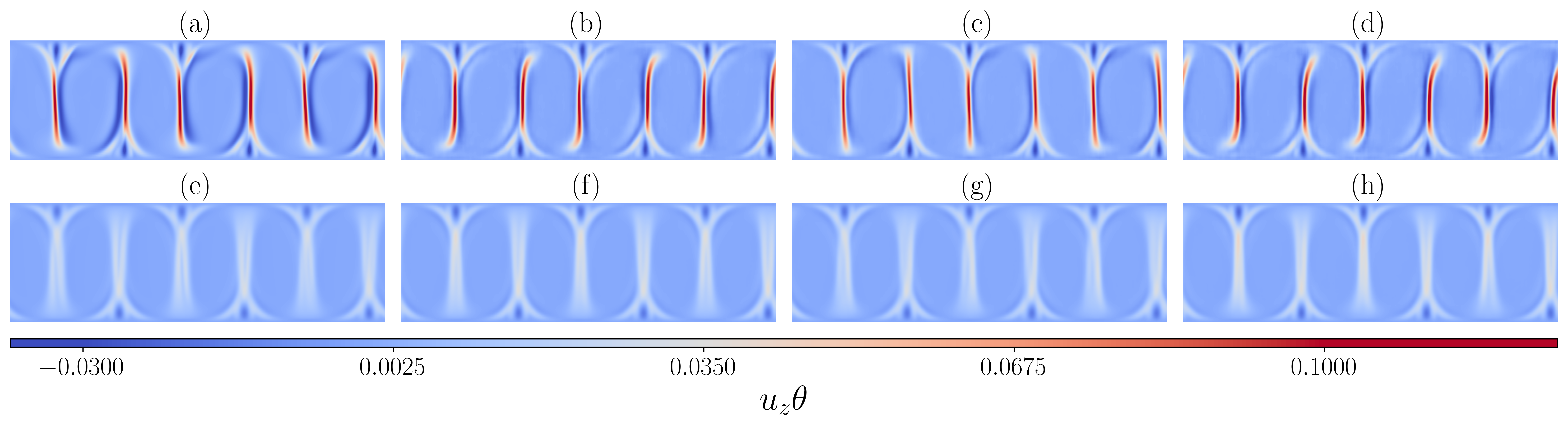}
\caption{Result from the blind test. (a)--(d) Instantaneous convective turbulent heat flux snapshots at a given time. (e)--(h) Mean convective turbulent heat flux field averaged over 900 snapshots. (a,e) DNS, i.e., ground truth. (b,f) CAE, i.e., ground truth for the ESN and GRU. (c,g) Prediction from the ESN with subsequent decoding. (d,h) Prediction from the GRU with subsequent decoding.}
\label{Figure04}
\end{figure*}
%------------------------------------------------------
\begin{figure}[htb!]
\centering
\includegraphics[width=0.73\linewidth]{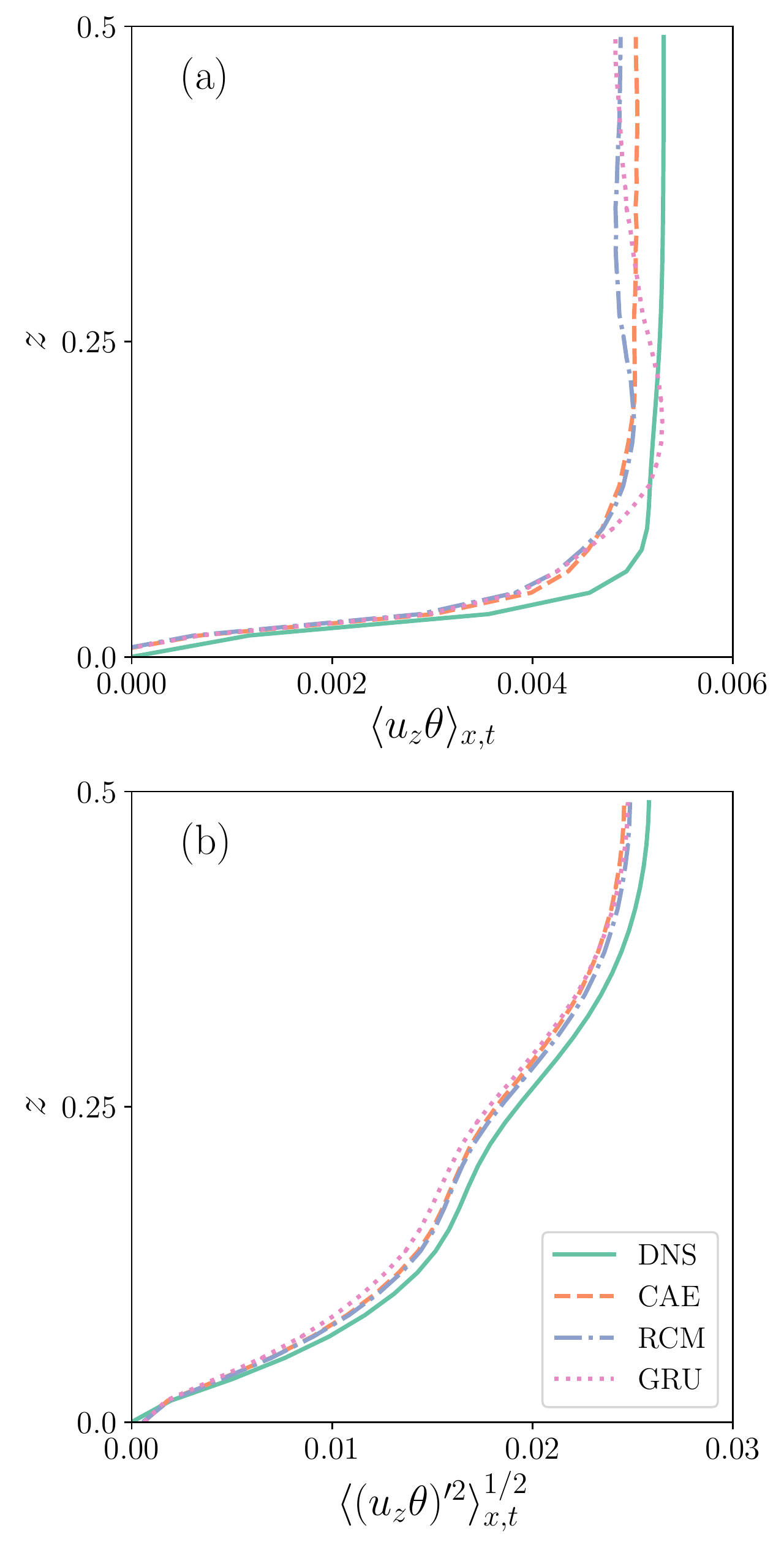}
\caption{Comparison mean convective turbulent heat flux and fluctuation over the half the cell height. Due to the top-down symmetry in RBC, we took an additional mean over both halves of the layer.  (a) Mean convective turbulent heat flux profile, and (b) Convective turbulent heat flux fluctuation profile. Linestyles in the legend hold for both panels.}  \label{Figure05}
\end{figure}
%------------------------------------------------------

\subsection{Convective heat flux fields and mean profiles}
With two levels of machine learning models combined, both of which have reached low errors in training and cross-validation, it is necessary now to assess their generalization ability by a comparison with unseen test data. These are taken out of the 900 remaining snapshots. Thereby, the ground truth DNS data are fed into the encoder, which generates lower-dimensional compressed data vectors of dimension $l=40$, see again Sec. III A. These latent vectors are then used as an initial condition to start the autonomous prediction by either the ESN or the GRU for the next 900 time steps. Afterwards, all 900 snapshots are passed through the decoder to reconstruct the actual flow field. Thereby, we evaluated two autoregressive prediction models, the ESN and the GRU. To evaluate the CAE itself, we fed all 900 snapshots into the encoder and and then directly reconstructed them via the decoder, thereby analyzing the reconstruction loss. 

Figure \ref{Figure04} depicts the qualitative comparison between the DNS results, the reconstruction by the CAE and the autonomous prediction by the ESN and GRU. A good agreement can be observed especially for the high-magnitude plume regions and the general flow structure in the form of circulating convection rolls, refer to Figs.~\ref{Figure04}(a)--(d). In the former regions the convective heat flux is locally largest, caused by sinking colder fluid or rising hotter fluid. In a 3d convection case, these regions would form a dynamically evolving skeleton as shown and analysed in Fonda et al.\cite{fonda2019deep}  

When looking at the instantaneous and time-averaged fields in Fig.~\ref{Figure04}, small deviations can be noticed due to the highly-reduced dimensionality of the latent space with $l=40$. This is an expected error for such kind of data compression. Nevertheless, one can conclude that the reconstructed fields agree qualitatively and even quantitatively fairly well with the ground truth. Unlike in cases, where a POD has been applied for data reduction \cite{Pandey2020a,Heyder2021}, the mean fields are not separated here; these fields were thus also dependent upon the ML predictions. 

A deviation of the mean profiles of the convective heat flux, $\langle u_z\theta(z)\rangle_{x,t}$, from the ground truth is seen in Fig.~\ref{Figure05}(a) away from the wall. This deviation is situated above the thermal boundary layer in the plume mixing zone for the high Prandtl number of ${\rm Pr}=7$ that is chosen here. In this region the dynamics is characterized by bursting thermal plumes that detach randomly at different positions from the wall and get dispersed by the turbulence. Clearly, the reproduction of this intermittent time dynamics of the convective turbulent transport is most challenging for the ML algorithms. 

Figure~\ref{Figure05}(b) shows fairly well overlapping profiles for the fluctuations of the convective heat flux for all 3 cases that remain close to the ground truth. Therefore 
\begin{equation}
  (u_z\theta)^{\prime}=u_z\theta-\langle u_z\theta(z)\rangle_{x,t}\,,
  \label{fluc}
\end{equation}
is defined. We have also quantified the loss of information of the CAE application which is based on the integrated convective flux which is given by
\begin{equation}
    \Phi=\int_A \langle (u_z\theta)^{\prime\,2}\rangle_t \;dA\,.
\end{equation}
The loss follows by 
\begin{equation}
    L_i=\frac{|\Phi_{i}-\Phi_{\rm DNS}|}{\Phi_{\rm DNS}}\,,
\end{equation}
with $i=\{{\rm CAE, ESN, GRU}\}$. The results are $L_{\rm CAE}=5.3\%$, $L_{\rm ESN}=4.2\%$, and $L_{\rm GRU}=6.7\%$.

\subsection{Probability density function of convective heat flux}
%------------------------------------------------------
\begin{figure*}[htb]
\centering
\includegraphics[width=0.9\linewidth]{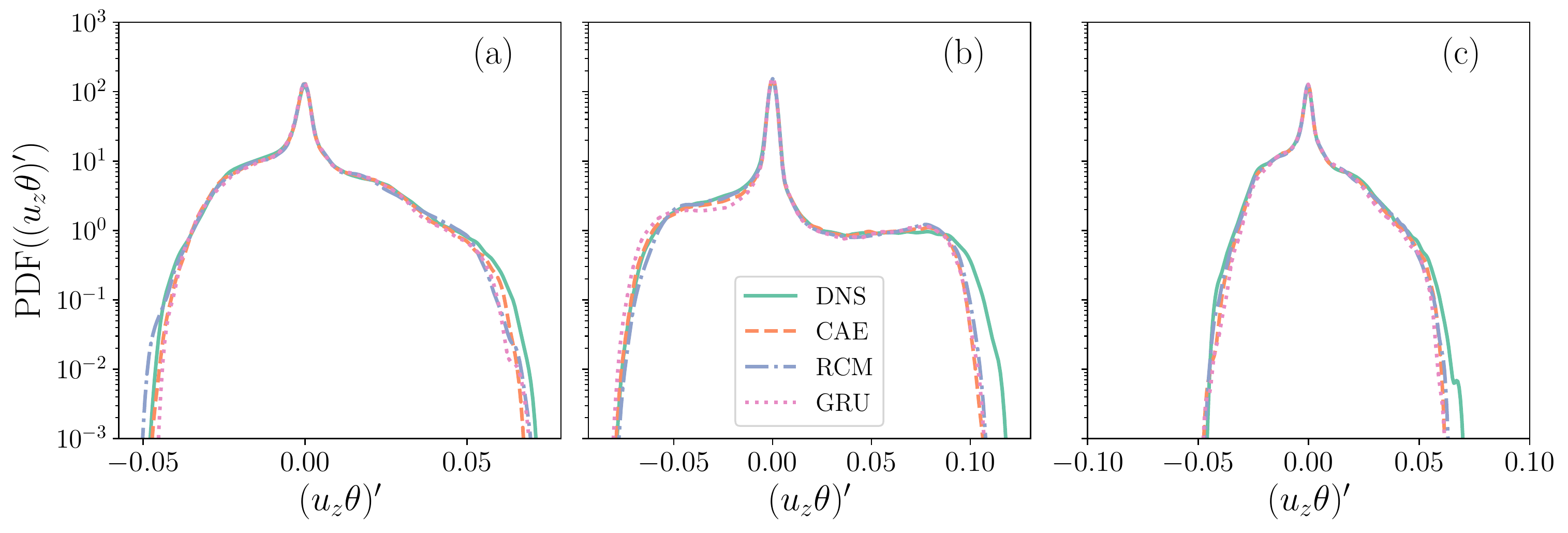}
\caption{Comparison of probability density functions (PDFs) for convective turbulent heat flux at 3 different locations in the wall normal direction $z$. (a) $z=0.16$. (b) $z=0.50$. (c) $z=0.84$. The legend holds for all three panels.} 
\label{Figure06}
\end{figure*} 
%------------------------------------------------------
\begin{figure}[htb]
\centering
\includegraphics[width=0.8\linewidth]{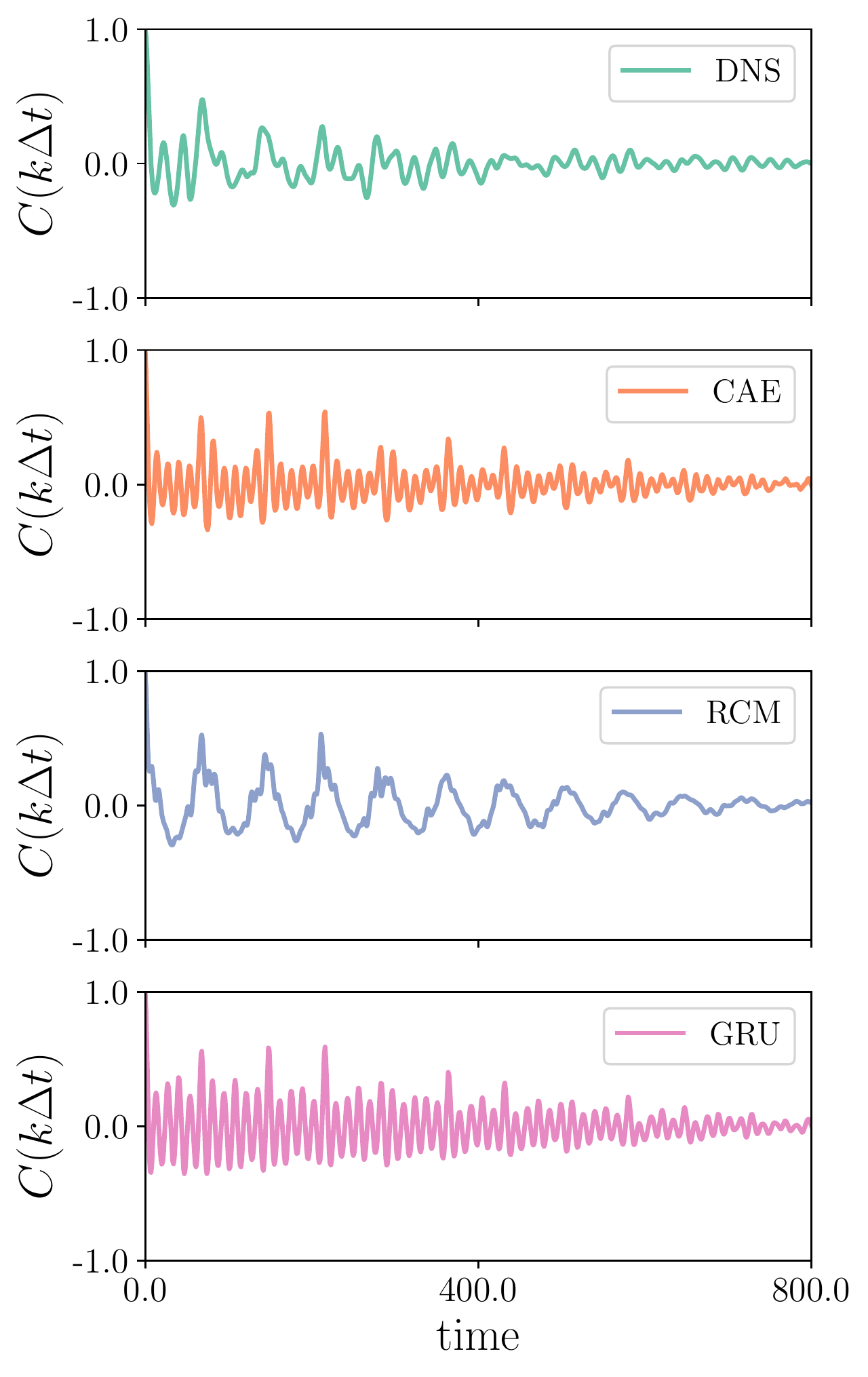}
\caption{Evolution of autocorrelation function $C(k\Delta t)$ for the convective turbulent heat flux $j_{\rm conv}$ at the middle line of the convection domain versus time $k\Delta t$ with $k=0,1,2,..$. The definition is given by eq. \eqref{corr}.}
\label{Figure07}
\end{figure} 
%------------------------------------------------------

Figure \ref{Figure06} illustrates a further comparison of the two ML algorithms with the original DNS results. The  probability density function (PDF) of convective turbulent heat flux $(u_z\theta)^{\prime}$, see eq. \eqref{fluc}, contains the full statistical information of the fluctuations at all orders. It is extracted here at three different locations in the channel. The PDFs of all methods show a good overlap around the mean. Deviations are observed in the tails, particularly in the positive tails where the most intense rising and falling plume events appear. The differences between pure CAE application and the combination with the ESN or GRU remain small. A physically important property of this PDF is that has to be skewed to positive values since heat is transported from the bottom to the top on average. This property is reproduced well by all our ML methods. The pronounced positive tails correspond physically to both, rising warmer-than-average and falling colder-than average thermal plumes.\cite{Shishkina2008,Emran2012}    

Figure~\ref{Figure07} depicts the temporal autocorrelation functions of the local convective heat flux, see eq. \eqref{conv1}, which were extracted from the middle of the convection domain at $z=1/2$. The autocorrelation is defined by 
\begin{equation}
    C(k\Delta t)=\dfrac{\sum_{k=0}^M\langle j_{\rm conv}(t)j_{\rm conv}(t+k\Delta t)\rangle_{x,z=1/2,t}}{\langle j_{\rm conv}^2\rangle_{x,z=1/2,t}}\,,  
    \label{corr}
\end{equation}
where $\Delta t=0.125$ is the time interval between subsequent output snapshots given in free-fall time units $H/U_f$. All algorithms, CAE, ESN, and GRU follow the typical trend where the autocorrelation decays gradually to zero with oscillations about the zero axis. The results for the DNS and the CAE show a qualitatively similar variation including the oscillation frequency of fluctuations and the amplitude. This is not the case for the ESN and GRU cases. The corresponding autocorrelations are found to oscillate at a much higher frequency. Both, ESN and GRU show also a slight mismatch regarding the amplitude of the local maxima and minima. This specific analysis demonstrates thus the limitations of the combined CAE-RNN application. The reproduction of the time correlations in the latent space is challenging.

%------------------------------------------------------
\begin{figure}[htb]
\centering
\includegraphics[width=0.8\linewidth]{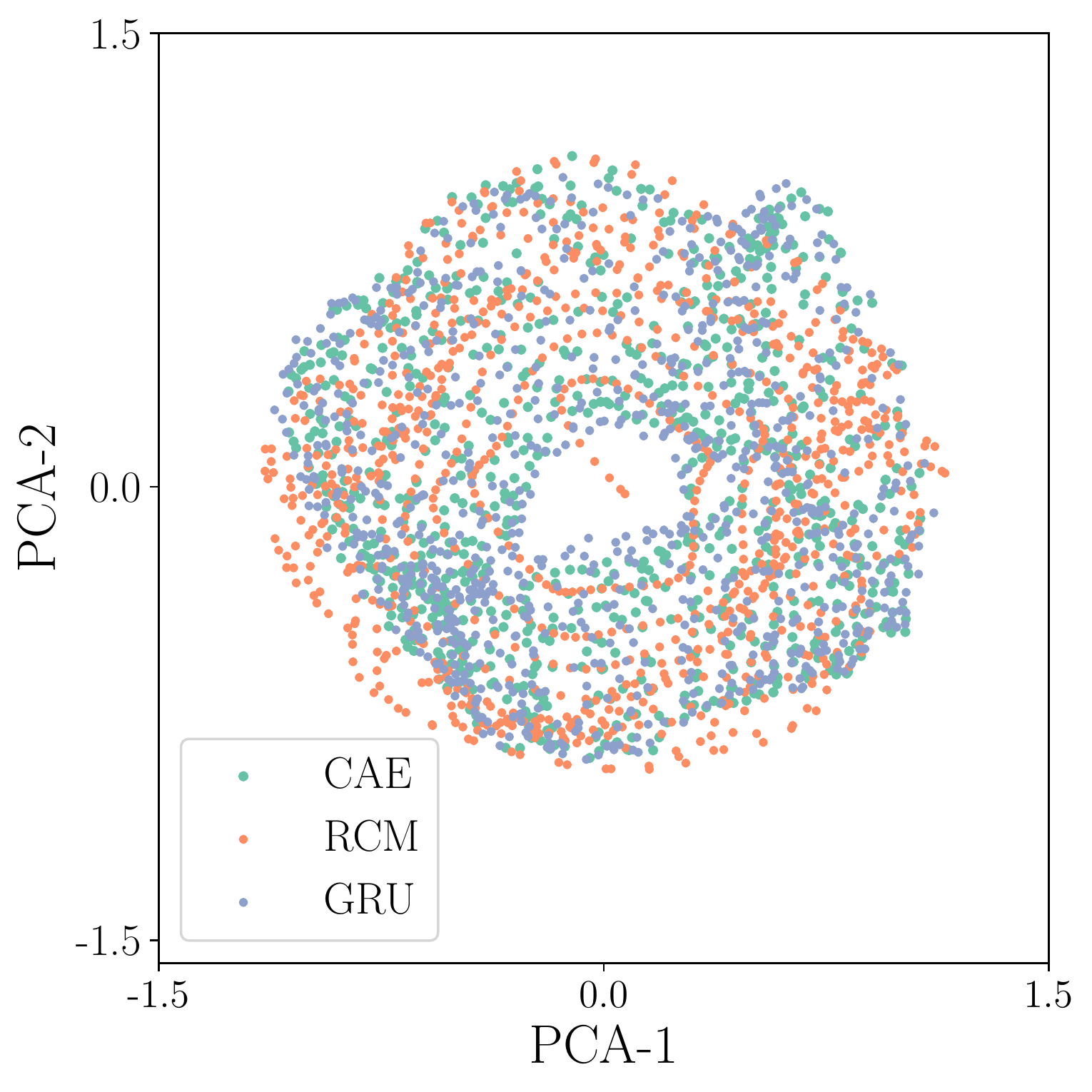}
\caption{Scatter plot visualization of the principal components obtained by a principal component analysis which was applied to the 40 modes in the latent space. The two primary components are denoted by PCA-1 and PCA-2.}
\label{Figure08}
\end{figure} 
%------------------------------------------------------

\subsection{Individually extracted modes and noise resilience}
%------------------------------------------------------
\begin{figure*}[htb]
\centering
\includegraphics[width=\textwidth]{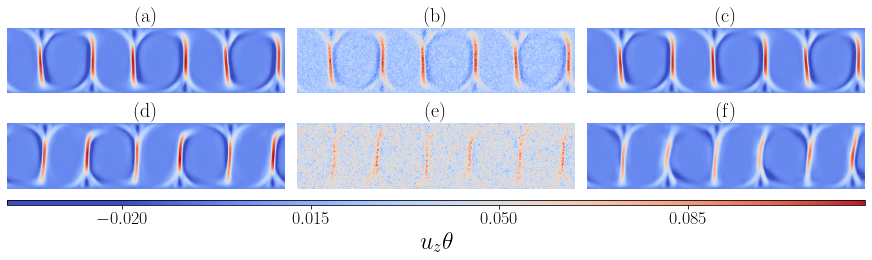}
\caption{Noise resilience of the CAE tested by means of contours of 2 individual temperature snapshots. (a,d) Prediction by the CAE for original temperature field without noise. (b,e) Original temperature field with added random noise. In panel (b), the standard deviation is $\sigma=0.025$, in panel (e) the distribution is broader with $\sigma=0.05$. (c,f) Prediction from the CAE for the corresponding noisy fields.}
\label{Figure09}
\end{figure*} 
%------------------------------------------------------

After directly comparing the flow fields with the ML prediction, we further verified the evolution of individual modes from the ESN evolution. This was done with the help of a principal component analysis (PCA). The PCA algorithm enables a further down-scaling from 40 modes in the latent space to 2 principal modes which carry more than 65\% of variance. Figure~\ref{Figure08} shows a scatter plot in 2D plane for comparison. We display the result of the CAE, i.e., ground truth for the subsequent RNN application together with the outputs of ESN and GRU for unseen test data. All data points are clustered with a similar range. As can be observed from the figure, the scatter plots overlap in the same region which warrants the learning and generalization ability of the ESN and GRU when compared to the CAE for the present dynamical system at hand.
%------------------------------------------------------
\begin{figure}[!htbp]
\centering
\includegraphics[width=0.8\linewidth]{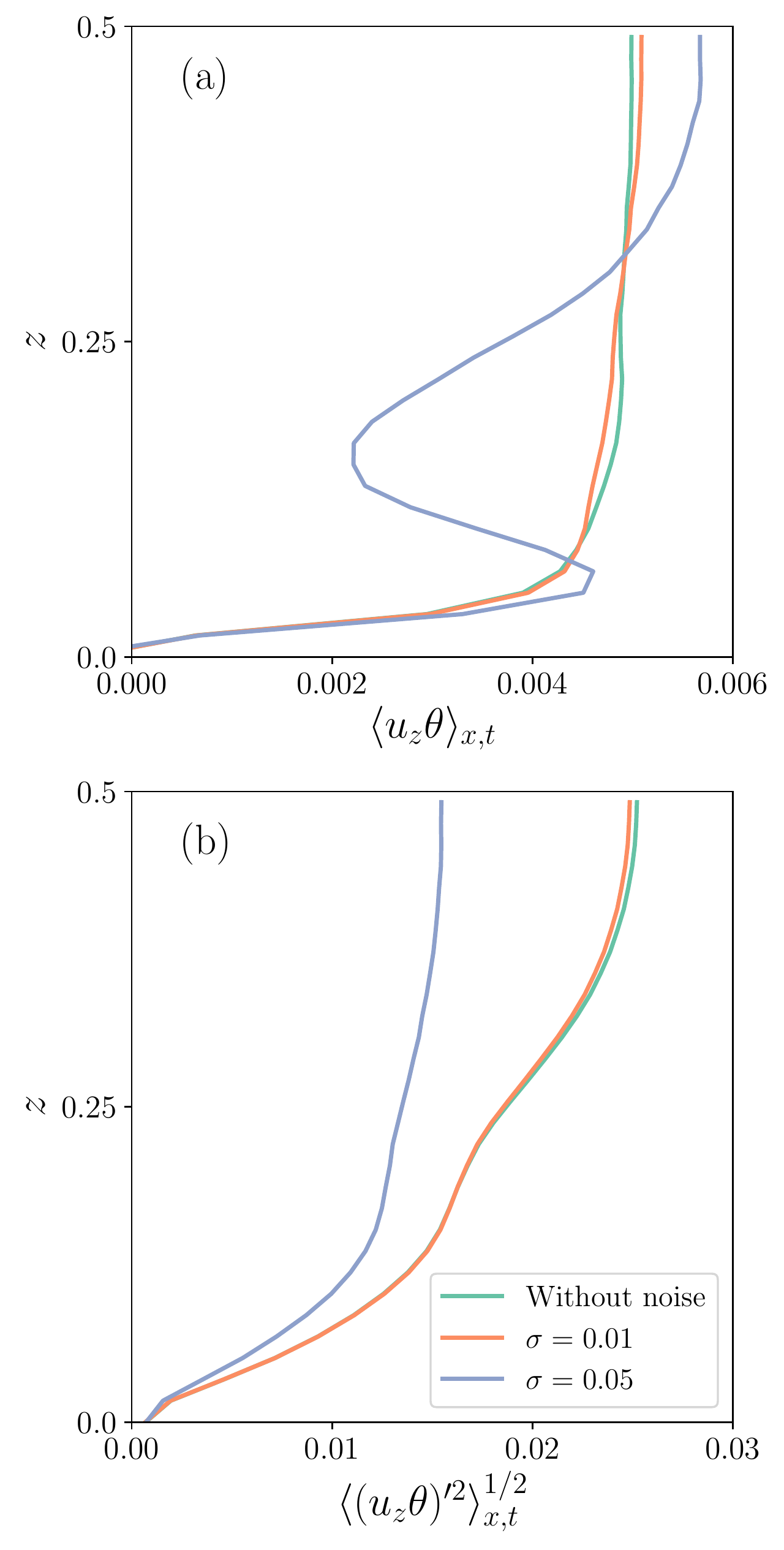}
\caption{Comparison of (a) the mean and (b) the fluctuation profiles over the cell height for 2 different level of random noise. Data are compared with the noise-free run.}
\label{Figure10}
\end{figure} 
%------------------------------------------------------

Finally, we examined the robustness of the CAE against noise. Robustness is desired especially when one wants to use a CAE in an experimental facility or for a simulation model with parametrizations and closures.  The objective of this step is to evaluate how the trained CAE will be affected by noise at the small scale. We investigated two distinct noise levels sampled from a random distribution with a zero mean and two different standard deviations $\sigma$. Figure~\ref{Figure09} shows that the CAE is affected by the noise. We apply the analysis to a temperature field snapshot here. However, at the lower noise level these effects remain subdominant and the CAE can filter them out (cf. Fig.~\ref{Figure09}a,c).  while a higher noise level is beyond the filtering abilities of the CAE (cp. Fig.~\ref{Figure09}d,f). 

Figure~\ref{Figure10} substantiates this finding. Small levels of noise leave the mean  and fluctuation profiles nearly unchanged. Higher levels of additive noise lead to stronger deviations in the low-order statistics as demonstrated in both panels of the figure. 

\section{Conclusions and outlook}\label{sec:conclusions}
In the present work, we have investigated two recurrent machine learning (ML) algorithms that model and forecast the local convective heat flux -- the central quantity for the characterization of the mean turbulent heat transfer from the bottom to the top -- in a two-dimensional turbulent Rayleigh-B\'{e}nard flow directly. This implies that this derived property, which is the product of the temperature and the vertical velocity component, together with its low-order statistics is modeled without applying the underlying highly nonlinear Boussinesq equations of motion. 

In both models, a convolutional autoencoder (CAE) is applied first to reduce the high-dimensional data records which are obtained from direct numerical simulations of the turbulent flow in a low-dimensional latent space. The dynamics in the latent space is advanced by means of either an echo state network (ESN), one implementation of a reservoir computing model, or a gated recurrent unit (GRU). We find that both ML algorithms performed well and are able to reproduce mean profiles of the convective heat flux and its fluctuations fairly well. This includes even the reproduction of the whole probability density function. It is furthermore tested how resilient the models are with respect to a small amounts of added noise. Our investigation demonstrate that the model is robust to such a noise with a small standard deviation of $\sigma\lesssim 0.02$ using the advantages of the CAE architecture for data reduction. The latter is often used for de-noising in image analysis.        

Typical fluid dynamics simulation and experiments inherently contain a large number of high-dimensional data records which make them unsuitable for a direct input into a machine learning algorithm. A reduction step, as applied here, is consequently necessary. Our CAE algorithm can be used on the fly (while the DNS is running and writing data records) and does not require the complete turbulence data set at the beginning of the reduction, as for example in case of a POD snapshot algorithm. Therefore, we presented a two-level neural network architecture which can reduce the dimensionality of data by using a non-linear convolutional autoencoder and the latent vector can be autoregressively predicted by a recurrent network. Due to a large number of hyperparameters in the ESN case, we took a Bayesian optimization, which enables an efficient search of optimized hyperparameters by a relatively small number of iteration steps in comparison to a conventional grid search. 

A potential application field of our CAE-RNN approach could be the modeling of mesoscale convective fluxes of heat (or moisture and  salinity) in global circulation models of the atmosphere and ocean. These models are typically built on coarse computational grids that span the globe and require parametrizations of locally strongly varying unresolved fluxes.\cite{Zanna2020,Bony2020} The developed model provides a dynamical ROM that delivers the low-order statistics of the turbulent transport in convection flows.      

The extension to three-dimensional data records is required and possible with the given tools. It will however face new additional challenges. Three-dimensional data records will require higher-dimensional latent spaces and deeper networks for both, encoder and decoder. This suggests a possible decomposition of the weight matrix of the corresponding convolutional networks into matrix product states which have been successfully used in the solution of problems in quantum many-particle dynamics.\cite{Orus2019} These investigations are currently in progress and will be reported elsewhere.    

\begin{acknowledgments}
The work is supported by the Deutsche Forschungsgemeinschaft with Grant No. SCHU 1410/30-1 and in parts by the project ``DeepTurb -- Deep Learning in and of Turbulence" which is funded by the Carl Zeiss Foundation. The authors gratefully acknowledge the Gauss Centre for Supercomputing e.V. (www.gauss-centre.eu) for funding this project by providing computing time through the John von Neumann Institute for Computing (NIC) on the GCS Supercomputer JUWELS at J\"ulich. 
\end{acknowledgments}

\section*{Data Availability Statement}
The data that support the findings of this study are available from the corresponding author upon reasonable request.

\appendix
\section{Convolutional Autoencoder Architecture}
The following Table \ref{tab:cae} details the architecture of the encoder-decoder network that was used in this work.
\begin{table*}[htb]
 \begin{tabular}{l c c l c} 
\hline\hline
\multicolumn{2}{c}{Encoder} & $\quad\quad\quad\quad\quad$ & \multicolumn{2}{c}{Decoder} \\
%\hline                    
Layer & Output size & & Layer & Output size \cr 
\hline
 Encoder input & $320\times 60 \times 1$ & & Decoder input & $5\times 1 \times 8$ \cr 
 2D Conv-E1 & $320\times 60 \times 256$ & &2D Conv-D1 & $5\times 1 \times 8$ \cr 
 Max Pool-E1 & $160\times 30 \times 256$ & & 2D Upsamp-D1 & $10\times 2 \times 8$ \cr 
 2D Conv-E2 & $160\times 28 \times 128$ & & 2D Conv-D2 & $10\times 2 \times 32$ \cr 
 Max Pool-E2 & $80\times 15 \times 128$ & & 2D Upsamp-D2 & $20\times 4 \times 32$ \cr 
2D Conv-E3 & $80\times 15 \times 64$ & & 2D Conv-D3 & $20\times 4 \times 32$ \cr 
Max Pool-E3 & $40\times 8 \times 64$ & & 2D Upsamp-D3 & $40\times 8 \times 32$ \cr 
 2D Conv-E4 & $40\times 8 \times 32$ & & 2D Conv-D4 & $40\times 8 \times 64$ \cr 
 Max Pool-E4 & $20\times 4 \times 32$ & & 2D Upsamp-D4 & $80\times 16 \times 64$ \cr 
  2D Conv-E5 & $20\times 4 \times 32$ & & 2D Conv-D5 & $80\times 16 \times 128$ \cr 
 Max Pool-E5& $10\times 2\times 32$ & & 2D Upsamp-D5 & $160\times 32 \times 128$ \cr 
  2D Conv-E6 & $10\times 2 \times 38$ & & 2D Conv-D6 & $160\times 32 \times 256$ \cr 
 Max Pool-E6 & $5\times 1\times 8$ & & 2D Upsamp-D6 & $320\times 64 \times 256$ \cr 
  &  &  & Output with 2D Conv & $320\times 64 \times 1$ \cr 
  & &  & Output with Cropping & $320\times 60 \times 1$ \cr 
\hline\hline
\end{tabular}
\caption{Detailed structure of the convolutional autoenocoder. The table summarizes the encoder and decoder architectures (Conv = convolution, Max Pool = max pooling, Upsamp = upsampling). Symbol E2 denotes for example encoder hidden layer No. 2.}
\label{tab:cae}
\end{table*}

\section{Hyperparameter tuning by Bayesian optimization}
The training of an ML algorithm relies on a cost function which depends on different hyperparameters ${\bm x}=(x_1,...,x_n)$. In the ESN case, the hyperparameter vector consists of $(N,D,\rho,\alpha,\beta)$.  Grid and random search procedures are often used and they proved to provide a (near) optimum solution.\cite{bergstra2012random} As a downside, these methods are based on a parameter space which is predefined in the form of a multi-dimensional  grid of parameter vectors. A more favorable alternative is the Bayesian optimization (BO), a global optimization method that automatically finds the optimal hyperparameter vector
\begin{equation} 
{\bm x}^{\bigstar} = \mathop{\rm argmax}_{{\bm x} \in \mathcal{X}} f({\bm x})\,,
\label{eq:argmax} 
\end{equation}
by a relatively small number of iterations.\cite{feurer2019hyperparameter} As the name suggests, BO utilizes the Bayes rule,
\begin{equation}
    p(f({\bm x})|{\bm x})\sim p({\bm x}|f({\bm x}))p(f)\,.
\end{equation}
The a-posteriori probability of a hyperparameter model $f({\bm x})=(f(x_1),...,f(x_n))$ given the hyperparameters ${\bm x}$ is similar to the likelihood of ${\bm x}$ given $f$, denoted as $p({\bm x}|f)$, and the a-priori probability $p(f)$. Here, $p(f)$ contains our obtained knowledge from prior iterations which is not discarded. Two main ingredients are necessary: 

(1) BO typically utilizes a Gaussian process (GP) to model $p(f)$ which is characterized by a mean $\mu({\bm x})$ and a covariance matrix $k(x_i,x_j)$. Here, we use a Mat\'{e}rn kernel for the covariance matrix which is given by\cite{genton2001classes}  
\begin{equation*} 
k(x_i, x_j) = \frac{1}{\Gamma(\nu)2^{\nu-1}}\left[\frac{\sqrt{2\nu}}{l} d(x_i,x_j) \right]^\nu K_\nu\left(\frac{\sqrt{2\nu}}{l} d(x_i,x_j )\right).
\label{eq:matern}
\end{equation*}
In this equation, $d(\cdot,\cdot)$ is the Euclidean distance, $K_{\nu}$ is a modified Bessel function of the second kind, and $\Gamma$ is the gamma function. The parameter $\nu$, which controls the smoothness of the learned function, is set to $\nu=1.5$. 

(2) BO needs furthermore an acquisition function to determine the hyperparameter vectors that are going to be evaluated by $f$ in the next iteration. This is a trade-off between the exploration i.e., to sample at a high uncertainty region and exploitation, i.e., querying a high mean region. In this work, we have used upper confidence bound (UCB) as an acquisition function.\cite{srinivas2009gaussian} UCB at iteration step $t$ is then given by 
\begin{align} 
{\bm x}_{*}= \operatorname*{arg\,max}_{\bm x} (\mu_{t}({\bm x})+ \kappa\sigma_t({\bm x}))\,,
\label{ucb}
\end{align}
where $\mu_t$ is the mean and $\sigma_t$ the standard deviation. Here, $\kappa$ is a UCB model coefficient which is provided in the main text when UCB is applied.
%\nocite{*}
\bibliography{references}% Produces the bibliography via BibTeX.

\end{document}